\documentclass[prb]{revtex4}

\usepackage{mathptm}    
\usepackage{dcolumn}                    
\usepackage{bm}                         

\usepackage{graphicx}           

\begin{document}

\preprint{LA-UR-03-5330}

\title{Lattice Dynamics and the High Pressure Equation of State of 
         Au}

\author{C. W. Greeff and M. J. Graf}

\affiliation{Los Alamos National Laboratory, Los Alamos, NM 87545, USA}

\date{\today}

\begin{abstract}

Elastic constants and zone-boundary phonon frequencies of 
gold are calculated 
by total energy electronic structure methods to twofold compression.
A generalized force constant model is used to interpolate throughout the 
Brillouin zone and evaluate moments of the phonon distribution. The
moments are used to calculate the volume dependence of the Gr\"uneisen
parameter in the fcc solid. Using these results with ultrasonic and 
shock data, we formulate
the complete free energy for solid Au. This free energy 
is given as a set of closed form expressions, which are valid to compressions
of at least $V/V_0 = 0.65$ and temperatures up to melting. Beyond this
density, the Hugoniot enters the solid-liquid mixed phase region. 
Effects of shock melting on the Hugoniot are discussed within an 
approximate model. We compare with proposed standards for the equation of state
to pressures of $\sim 200$ GPa. Our result for the room temperature
isotherm is in very good agreement with an earlier standard of Heinz and
Jeanloz.

\end{abstract}

\maketitle

\section{Introduction}

The elastic constants, phonon frequencies, and equation of state (EOS)
are fundamental properties of matter. The values of these
parameters under compression find application in 
geophysics~\cite{manghnani} and in
the prediction and interpretation of processes of dynamic 
compression.~\cite{mcqueen,bgp} For many materials, especially elemental
metals, the principal Hugoniot curves, the set of states accessible via
a single shock from ambient conditions, have been 
measured.~\cite{mcqueen,llnl,lasl,altshuler} Since the pressure, density,
and internal energy are known along the Hugoniot from the 
jump conditions,\cite{mcqueen} 
this data is an important baseline for high pressure
equations of state. The off-Hugoniot EOS is needed for
the prediction of processes involving more complicated loading paths,
such as multiple shocks or shock and release. \cite{nellis} 
Extrapolation from the Hugoniot has also been used to establish pressure
standards for static high pressure experiments, whose room temperature 
isotherms are regarded as known. \cite{mcqueen,jamieson,mbss}
Au has been used as a standard in this way, and has been
used to calibrate the ruby $R_1$ line as a secondary 
standard.\cite{mbss,mxb,bell_sccm}
Recent studies have called into question the accuracy of the gold pressure standard.
Akahama {\it et al.}\cite{akahama} compressed Au and Pt in the same cell and found
two Au standards\cite{jamieson,heinz} to give pressures lower than Pt by 20
and 15 GPA, respectively, at 150 GPa. Shim {et al.},\cite{shim} on the other hand,
propose a new EOS for Au that gives pressures still lower than either of these
earlier standards.

To relate the Hugoniot to the room temperature isotherm requires information
on the Gr\"uneisen parameter $\gamma = V \left( \partial P/\partial E \right)_V$
and the specific heat $C_V$.\cite{mcqueen} These are dominated 
by lattice vibrations in
the regimes under consideration here. With increasing compression,
the separation between the Hugoniot pressure and the room temperature
pressure increases, resulting in greater dependence of the inferred
room temperature isotherm on $\gamma$. Because
$\gamma$ is not easily measured at high pressure, this introduces a 
non-negligible source of 
uncertainty in the pressure standards. In most cases, a model assumption
of the form $\gamma(V) = \gamma(V_0) (V/V_0)^q$ has been used. The
specific value $q=1$ has been used often in shock work.\cite{mcqueen,jamieson}
However,
a power law dependence of $\gamma$ on $V$ is qualitatively incorrect
at high compression, and extrapolation on this basis is inherently limited.

In principle, the various components of the EOS -- the static lattice
energy, the vibrational free energy, and the electronic excitation free
energy -- can be evaluated from electronic structure theory. Practical
calculations based on approximate density functional theories typically
have errors of several percent in the density at zero pressure, and 
10 \% or more in the bulk modulus. \cite{khein} These are unacceptably
large errors for the purpose of high accuracy equations of state.
These properties can be measured accurately, and are mainly determined 
by the cold energy curve, which
is the largest contribution to the EOS in the regime considered here.
On the other hand, we find that {\it ab initio} electronic structure 
calculations are capable of obtaining phonon frequencies of 
sufficient accuracy to strongly constrain the volume dependence
of the Gr\"uneisen parameter $\gamma$. We therefore propose that 
the most accurate EOS in
the solid is obtained by combining an empirical cold energy with
{\em ab initio} lattice vibration and electronic excitation free energies.
This is analogous to procedures that have been used to reduce
shock Hugoniot data and derive room temperature standards, but
the present analysis gives a strong physical foundation for the volume
dependence of $\gamma$, allowing confidence in the results at higher
compression. We are not aware of any evidence of solid-solid phase
transitions in Au, and we consider only the fcc solid. Recent 
calculations\cite{boettger_au}
indicate a transition to the hcp structure for $V/V_0 \alt 0.6$,
and we do not attempt to extend our solid EOS beyond this density.

This paper is arranged as follows. First we discuss lattice dynamics
and its connection to the EOS. We emphasize the importance of the classical limit
for defining the ion motion contribution to $\gamma$. 
We describe our procedures for calculating phonon frequencies 
and interpolating to the whole Brillouin zone. Next we present our results
for elastic and phonon properties of  Au, and our analysis
to obtain $\gamma^{\rm ion}(V)$. We then present the complete EOS for
solid  Au by giving a set of closed form expressions for the Helmholtz
free energy with parameters. The resulting room temperature isotherm
is compared with various proposed standards. Next we discuss
shock melting and its effect on the Hugoniot within an approximate model,
and finally give our conclusions.

\section{Lattice Dynamics and the Equation of State}

We write the Helmholtz free energy as
\begin{equation}
F(V,T) = \phi_0(V) + F^{\rm ion}(V,T) + F^{\rm el}(V,T)
\label{f_total}
\end{equation}
where $\phi_0$ is the static lattice energy, $F^{\rm ion}$ is the ion motion
free energy, and $F^{\rm el}$ is the electronic excitation free energy.
In our applications, $\phi_0$ gives the largest contribution to the
pressure. The ion motion free energy gives the dominant temperature
dependence of $P=-\left(\partial F/\partial V\right)_T$. The electronic
excitation term  $F^{\rm el}$ is
generally a small correction to the solid EOS, becoming important
for the liquid Hugoniot at several hundred GPa.

It is, therefore, imperative to have an accurate $F^{\rm ion}$ for calculating
the temperature dependence of $P$.
The quasi-harmonic approximation has been
found to have small errors in many cases.\cite{wallace_therm}
We have carried out
Monte Carlo simulations using an embedded atom model\cite{voter} of Cu
to investigate the importance of anharmonic corrections to
the ion free energy.
Here we define the anharmonic free energy to be the difference
between the true ion free energy and the quasi-harmonic approximation,
noting that the term has been
used differently\cite{anderson_au} by other authors.\cite{footnote01}
We intend to publish details of our Cu simulations elsewhere. In summary,
we find that up to 300 GPa on the melting curve, $ | P^{\rm anh} | < 0.85 $ GPa,
and is never more than 3\% of the {\it thermal} pressure $P(V,T)-P(V,0)$,
a very small correction to the total pressure.
Given the similarity of the bonding in Cu and Au, we expect these results
to be relevant for Au also.  Thus, in what follows we
neglect anharmonicity and use the quasi-harmonic approximation for $F^{\rm ion}$,
\begin{equation}
F^{\rm ion}(V,T) = \int_0^{\infty} d\omega \, g(\omega) \left[ \frac{1}{2} \hbar \omega
                       + k_B T \ln( 1 - e^{-\hbar \omega/k_B T}) \right] \, ,
\label{fqh}
\end{equation}
where we have introduced $g(\omega) = \frac{1}{3N} \sum_{\bf k} 
 \delta(\omega - \omega_{\bf k})$ ,
 the normalized phonon density of states.
The phonon frequencies $\omega_{\bf k}$ are
functions of volume only,
and the sum is over the $3N$ normal modes of the crystal.

Application of Eq.~(\ref{fqh}) requires the full phonon density of
states $g(\omega)$ at all volumes. While in principle this information
is available from our calculations, in practice the classical limit
dominates our EOS, allowing for a substantial simplification. In the
classical limit, which is the leading term in the high temperature
expansion of Eq.~(\ref{fqh}), the free energy is given by
\begin{equation}
F^{\rm cl}(V,T) = 3N k_B T  \ln\left(\frac{e^{-1/3} \hbar \omega_0}{k_B T} \right)   \, .
\label{f_cl}
\end{equation}
where we have introduced the moment
\begin{equation}
\omega_0  =  e^{1/3} \exp\left[ \int_0^{\infty} d\omega \, g(\omega) \ln\omega \right ]  \, .
\label{omega_0}
\end{equation}
Other moments
are conventionally defined as 
\begin{equation}
\omega_n  =    \left[ \frac{3+n}{3} \int_0^{\infty} d\omega \, g(\omega) \omega^n \right]^{1/n}
                 n \neq 0 , n > -3 \, ,
\label{omega_n}
\end{equation}
where the normalizations in Eqs.~(\ref{omega_0}) and 
(\ref{omega_n}) are chosen so that for a Debye spectrum, described by
\begin{equation}
g(\omega) = \frac{3}{\omega_D^3} \omega^2  \Theta(\omega_D - \omega) \, ,
\label{g_debeye}
\end{equation}
all $\omega_n$ are equal to $\omega_D$, the Debye frequency. We subsequently
give results for $\nu_n = \omega_n /(2 \pi)$ corresponding to the
experimental convention of giving frequencies $\nu$ in Hertz,
as opposed to angular frequencies, $\omega$.

\begin{figure}
\noindent
\includegraphics[scale=0.35]{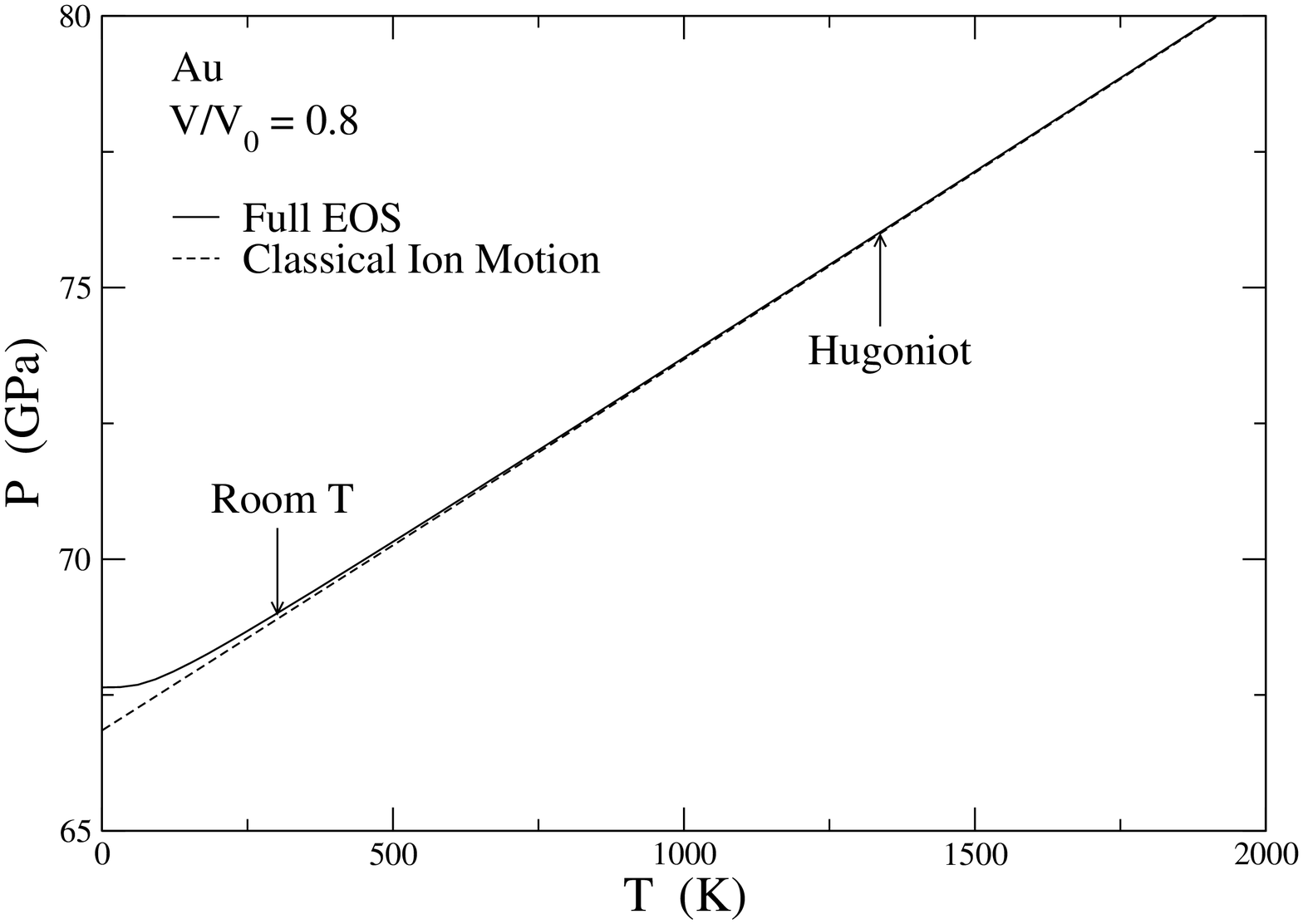}
\caption{Role of quantum ion motion in the EOS. Graph shows $P(T)$
         along an isochore $V/V_0=0.8$. Solid curve is the full EOS, dashed curve
         uses the classical limit of the ion free energy. Hugoniot temperature
         at the given density is 1340~K.}
\label{qcl_plot}
\end{figure}

In the classical limit, the ion pressure is linear in $T$
with $\left( \partial P^{\rm ion}/\partial T\right)_V
= \left(3N k_B /V\right) \gamma_0$, 
where $\gamma_0 = -\frac{d \ln \omega_0}{d \ln V}$, and the specific 
heat is constant, $C_V = 3 N k_B$.
The role of quantum ion motion in the EOS is illustrated
in Figure \ref{qcl_plot}, which shows
the pressure along an isochore $V/V_0=0.8$. The solid curve is our full EOS,
described below, and the dashed curve uses the classical limit of $F^{\rm ion}$
at all $T$. The arrows mark room temperature and the Hugoniot temperature,
which is 1340 K at the given density. The melting temperature at this density is
estimated to be 4900~K, beyond the range of the plot. It is clear that
the classical limit dominates even at room temperature. The Hugoniot
is well into the classical regime. At higher densities, the Hugoniot
is still further above the quantum regime. The largest quantum effect on the
pressure is at $T=0$, where the zero-point vibrations contribute
a pressure of 0.8 GPa at this density. The classical limit is clearly 
dominant for interpolating between the Hugoniot and room temperature.

We have found that an interpolation based on the Debye model
gives a very accurate approximation to the full quasi-harmonic free energy.
The Debye free energy is a special case of the quasi-harmonic
free energy, Eq. (\ref{fqh}). Inserting the Debye density of states,
Eq.~(\ref{g_debeye}), gives
\begin{equation}
F^{\rm D}(V,T) = N \left[ \frac{9}{8} \hbar \omega_D
                      + 3 k_B T \ln \left( 1 - e^{-\hbar \omega_D /k_B T} \right)
                      - k_B T D\left(\hbar \omega_D /k_B T\right) \right] \, ,
\label{f_debye}
\end{equation}
where
\begin{equation}
D(x) = \frac{3}{x^3} \int_0^x dz \frac{z^3}{e^z -1} \, .
\label{debye_fn}
\end{equation}
In light of the above remarks, it is important to capture the 
correct classical limit. This requires that we set
\begin{equation}
\omega_D(V) = \omega_0(V)  \, .
\label{wdw0}
\end{equation}
To emphasize the distinction between Eq.~(\ref{wdw0}) and the
standard definition of $\omega_D$ in terms of acoustic velocities, we
refer to the Debye free energy together with Eq.~(\ref{wdw0})
as the high-temperature Debye model. The high-temperature Debye
model gives the same results as the full quasi-harmonic free energy
in the classical regime, and in addition obeys the Nernst theorem
at low $T$. We have checked that 
at low $T$, the error in the pressure compared to the full quasi-harmonic 
approximation is entirely negligible.

Thus, we can simplify the specification of the lattice vibration free energy
from giving $g(\omega)$ at all $V$ to giving the single moment
$\omega_0$ at all $V$. This is advantageous for numerical applications.
It also allows us to express our full EOS in a few compact formulae  
so that it is generally accessible. These formulae are given below.

Calculating the moments $\omega_n$ requires the phonon frequencies for
all ${\bf k}$ in the Brillouin zone. Direct {\it ab initio} calculations
on a dense mesh in ${\bf k}$ would be quite expensive. As a result of 
another study\cite{graf_pdf} we found that the low order moments can often
be accurately obtained from short-ranged force constant models. In
particular, for  Au, the moment $\omega_0$ is converged to less 
than 1 \% with a 2nd neighbor (2NN) model. Therefore, our method is to 
calculate four zone boundary phonon frequencies corresponding to
the transverse and longitudinal modes at the $X$ and $L$ points. These
are calculated with standard frozen-phonon methods.
In addition the three elastic moduli are calculated using the method
described by S\"oderlind et al.\cite{soderlind_elastic} We fit these
results to a 2NN force constant model, which then allows  
the evaluation of $\omega_{\bf k}$ for arbitrary ${\bf k}$ for integration
over the Brillouin zone.

The electronic structure calculations used the full-potential 
LAPW code WIEN97.\cite{Wien97}
We used the LDA rather than the GGA, based on Boettger's finding\cite{boettger_au} 
that the LDA gives a better static lattice energy than the GGA for Au.
Some numerical parameters used in the calculations  were,
in atomic units:
muffin tin radius $r_{\rm MT} = 2.0$; plane wave
cut-off $r_{\rm MT} k_{\rm max} = 9.0$; cut-off for expansion of
density and potential $g_{\rm max} = 14$. For elastic modulus
calculations, Brillouin zone integrals
used special points corresponding to  $18^3$ points in the
full zone, with Gaussian smearing of the energies by 20 mRy.
The zone boundary phonons were found to be comparatively insensitive
to the ${\bf k}$-point mesh, and smaller meshes of $10^3$ points were used.
The $5p$, $5d$, and $6s$ shells were treated as valence states,
and  local orbital extensions\cite{singh_lo} were used in the $p$ and $d$
channels.

We used a generalized Born-von K\'arm\'an force model for our lattice-dynamical 
calculations, from which we were able to compute the phonon
dispersions in the entire Brillouin zone. Since gold has a very simple
phonon dispersion, we employed only a model with first (1NN) and second 
(2NN) nearest-neighbor interatomic shells of atoms. In an fcc lattice
the 1NN and 2NN forces are determined by three and two independent
parameters, respectively (for more details see e.g. ref.~\onlinecite{graf_pdf}).
These five force constants were extracted by fitting simultaneously 
the phonon frequencies of gold at the $X$ and $L$ points of the Brillouin 
zone and the elastic constants C$_{11}$, C$_{12}$, and C$_{44}$ near 
the zone center ($\Gamma$ point). Thus we fit a total of seven independent 
data points.  We gave  equal
weight to the zone boundary phonons and elastic constants in our $\chi^2$-fit of the
phonon dispersions.
Although the 2NN Born-von K\'arm\'an force model is too simple to reproduce
all phonon frequencies within less than approximately 10\%, 
see Figure \ref{dispcurve},
it is sufficiently accurate to compute integrated quantities such as the 
phonon moments within less than approximately 3\%. More accurate
phonon dispersions can be obtained if needed, by computing {\it ab initio}
frequencies at half and quarter distances in the Brillouin zone and fitting
those to a 3NN or higher-order Born-von K\'arm\'an force model, or to phonon
models with interatomic pseudo-potentials.

\section{Elastic and Phonon Properties}

\begin{figure}
\noindent
\includegraphics[angle=270,scale=0.35]{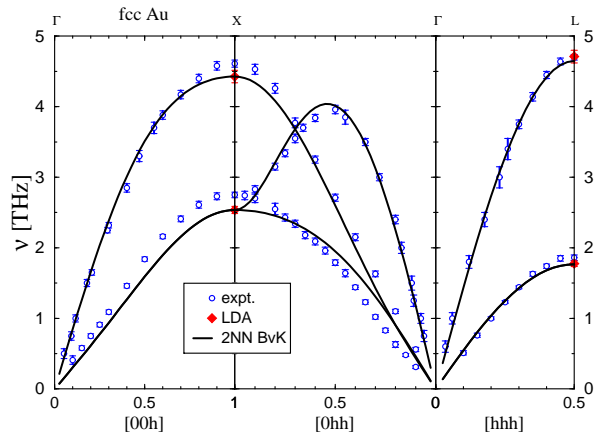}
\caption{(Color Online) Phonon dispersion for Au at room temperature and ambient pressure.
          Open circles are experimental data of 
         Lynn {\it et. al.}~\cite{Lynn1973}. Filled diamonds are LDA calculations 
           of zone boundary phonons at the $X$ and $L$ points. Solid curve 
           is interpolation based
           on fit of 2NN force constant model to the
           LDA zone-boundary phonons and elastic moduli.}
\label{dispcurve}
\end{figure}

Figure \ref{dispcurve} shows the calculated phonon dispersion curves for
Au at the density corresponding to ambient pressure and room temperature. 
The filled diamonds at the
$X$ and $L$ points are the LDA frozen phonon results. These, together with the 
calculated elastic moduli are used to obtain the force constants. The solid curves
are the interpolation to general wavevectors using the force constant model.
The open circles are the experimental data of Lynn {\it et al.}\cite{Lynn1973}
The force constant model is over-constrained, so it neither goes precisely through
the calculated zone boundary frequencies, nor has it exactly the calculated 
elastic moduli. The shapes of the dispersion curves
are simple enough for Au that they are generally well captured by the
2NN model. 

Table \ref{au_phonon_table} summarizes our results for the zone boundary phonons,
elastic moduli, and moments $\nu_n = \omega_n/2\pi$, as functions 
of volume. The reference
volume $V_0$ corresponds to $T= 298$ K and $P= 1 $~bar~$ = 100$~kPa. 
For Au, $V_0 = $
10.212~cm$^3$/mol or 114.43~$a_0^3$/atom. Also shown are the experimental data
at room temperature.\cite{Lynn1973,simmons} No attempt is made to account for the temperature
dependence beyond comparing at the correct volume. There is generally
good agreement between the calculated and experimental quantities. The
main exception is $C_{44}$, which is calculated substantially lower than
measured. Our calculation is in better agreement with an earlier 
LDA calculation.\cite{soderlind_elastic} The main result of the 
present calculation is the value of $\nu_0$, which is 
within 3\% of the measurement. The ratio $\nu_1/\nu_0$ is 1 for a Debye
spectrum. Our calculations give $\nu_1/\nu_0 = 1.03$ at ambient density
and 1.04 at twofold compression, so, by this measure, the departure
from a Debye spectrum is small and nearly constant with volume.

\begin{table}
\caption{Calculated phonon and elastic properties of Au. $V_0$ is the volume
at 298 K and atmospheric pressure and is 10.212 cm$^3$/mol for Au. Frequencies
$\nu$ given in THz and elastic moduli given in GPa.}
\label{au_phonon_table}
\begin{tabular}{lrddddddddd}
$V/V_0$  &  $\nu_{Xt}$  & \nu_{Xl}  &  \nu_{Lt} & \nu_{Ll}  &  C' & C_{44} & %
          B & \nu_0  &  \nu_1  &  \nu_2 \\
\hline
 1.1  &    1.76 &    3.17 &    1.33 &    3.37 &     9.2 &    11.1 &    91.0 &    2.53 &    2.60 &    2.68 \\
 1.0  &    2.54 &    4.43 &    1.77 &    4.71 &    13.8 &    27.4 &   172.0 &    3.53 &    3.64 &    3.75 \\
1.0/Expt.  & 2.75 & 4.61 & 1.86  & 4.70 & 14.6 & 41.5 & 167. & 3.65 & 3.75 & 3.86 \\ 
 0.9 &    3.45 &    5.91 &    2.30 &    6.32 &    20.3 &    55.0 &   304.4 &    4.71 &    4.86 &    5.00 \\
  0.8 &    4.55 &    7.71 &    2.95 &    8.40 &    26.8 &   112.3 &   527.3 &    6.16 &    6.35 &    6.54 \\
  0.7 &    5.94 &   10.05 &    3.76 &   11.19 &    37.4 &   221.7 &   928.3 &    8.03 &    8.30 &    8.57 \\
  0.6 &    7.76 &   13.18 &    4.78 &   15.23 &    67.0 &   443.5 &  1663.8 &   10.55 &   10.93 &   11.30 \\
  0.5 &   10.27 &   17.63 &    6.10 &   21.62 &   135.4 &   871.0 &  3097.4 &   13.99 &   14.58 &   15.13 \\
\end{tabular}
\end{table}

\begin{figure}
\noindent
\includegraphics[scale=0.35]{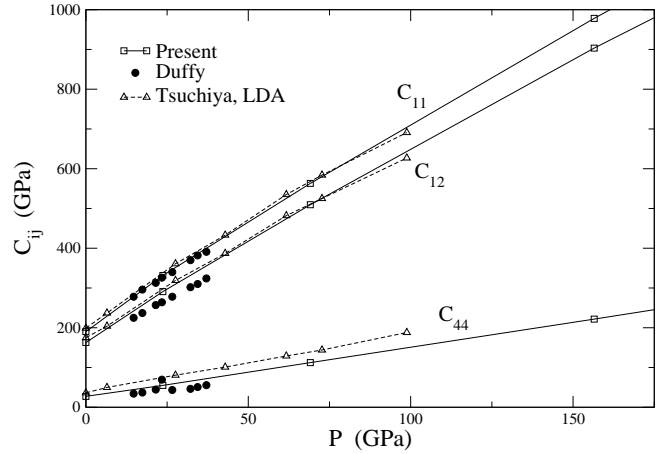}
\caption{Pressure dependence of the elastic moduli of Au. Open squares are
present LDA calculations with pressures from our EOS. Solid circles are 
experimental data from Duffy {\it et al.}\cite{duffy_cij}. Open triangles 
are previous LDA calculations by Tsuchiya et al. \cite{tsuchiya}}
\label{cij_plot}
\end{figure}

Measurements of elastic moduli were reported by Duffy {\it et al.}\cite{duffy_cij}
to $P= 37$ GPa, and {\it ab initio} calculations by Tsuchiya and 
Kawamura\cite{tsuchiya} to
100 GPa. Figure \ref{cij_plot} shows these results along with our calculations.
Our volumes are converted to room temperature pressures using the
EOS described below. The highest density in table \ref{au_phonon_table}
corresponds to $P \approx 780$ GPa, and the graph is restricted to lower 
pressures to highlight the comparison with these other works. Our calculations
of $C_{11}$ and $C_{12}$ are in good agreement with Tsuchiya and Kawamura,
while our $C_{44}$ is systematically lower, in somewhat better agreement
with the experiments. Our $C_{11}$ is in good agreement with the experiments
while $C_{12}$ is slightly high. It should be noted that the experimental
$C_{ij}$ depend on a model parameter $\alpha$ used in the analysis.\cite{duffy_cij}
The plotted points correspond to $\alpha =1$. It is encouraging that our
calculations give the correct trends for the pressure dependence of $C_{ij}$.

\begin{figure}
\noindent
\includegraphics[scale=0.45]{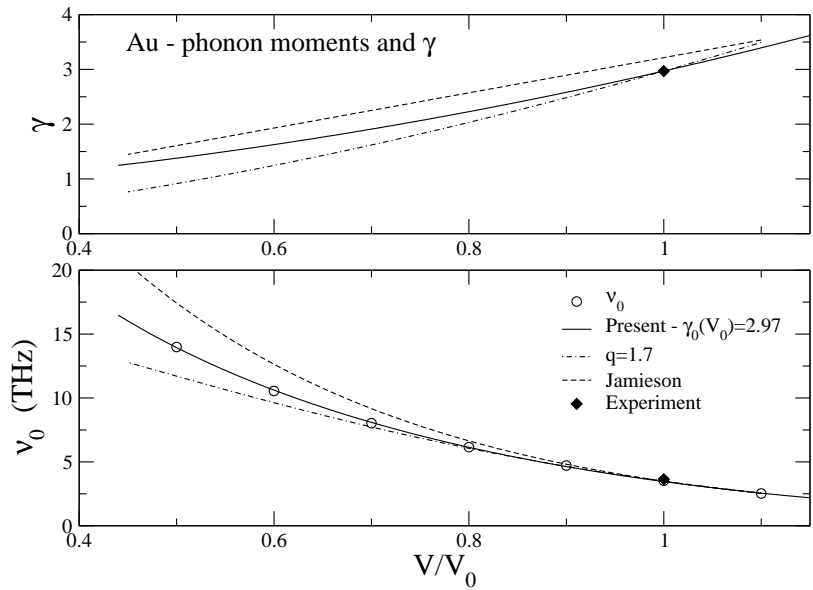}
\caption{Volume dependence of $\nu_0$, the log moment of the
         phonon frequencies, and $\gamma_0 = - d \ln \nu_0/d \ln V$ for Au.
         Lower graph is $\nu_0(V)$. Open symbols are calculated moments.
         Upper graph is resulting $\gamma_0(V)$. Solid curves are present
         results fitting Eq.~(\ref{gamma_of_v}) to the calculated moments
         with the constraint $\gamma_0(V_0) = 2.97$. Dash-dot curve shows the
         commonly used approximation $\gamma(V) = \gamma(V_0) (V/V_0)^q$
         for $q=1.7$.\cite{heinz} Dashed curve corresponds to the
         $\gamma$ of Jamieson {\it et al.}\cite{jamieson}, who used
         $q = 1$ and $\gamma(V_0)= 3.215$.}
\label{nu_gamma_plot}
\end{figure}

Having determined the moments $\nu_n$ as functions of volume, we need to 
interpolate and differentiate them to obtain thermodynamic functions. To do
this, we assumed the following functional form for $\gamma_n(V)$, which
has been used in creating many wide ranging equations of state,\cite{grizz}
\begin{equation}
\gamma_n(V) = \gamma^{\infty} + A_n \left( \frac{V}{V_0}\right) 
                + B_n \left( \frac{V}{V_0}\right)^2
\label{gamma_of_v}
\end{equation}
where $\gamma^{\infty}$ is the infinite density limit of $\gamma$, and
$A_n$ and $B_n$ are parameters. Alternatively, we can express $A_n$ and $B_n$
in terms of $q_n =d \ln(\gamma_n) /d \ln(V)$ and $\gamma_n(V_0)$ as
\begin{eqnarray}
A_n & = & \gamma_n(V_0) \left[ 2 - q_n(V_0) \right] - 2 \gamma^{\infty} 
  \nonumber \\
B_n & = & \gamma_n(V_0) \left[ q_n(V_0) -1 \right] + \gamma^{\infty}
\label{AB_gammaq}
\end{eqnarray}
Integrating $\gamma_n = - \frac{d \ln \nu_n}{d \ln V}$, we have
\begin{equation}
\nu_n(V) = \nu_n(V_0) \left( \frac{V}{V_0} \right)^{- \gamma^{\infty}}
         \exp \left\{ - A_n \left[ \left(\frac{V}{V_0} \right) -1 \right]
                    -\frac{B_n}{2} \left[\left(\frac{V}{V_0} \right)^2 -1 \right]
                 \right\}  \, .
\label{nu_of_v}
\end{equation}
By fitting this functional form to the calculated $\nu_0(V)$, we extract
the parameters giving $\gamma_0(V)$. In our fits, we keep $\gamma^{\infty}$
fixed. The value $\gamma^{\infty} = 2/3$ has been widely used,\cite{bushman}
although arguments have been made for $\gamma^{\infty} = 1/2$.\cite{lb_gamma}
For our application and density range, we have found that the quality 
of the fit and the 
resulting $\gamma_0(V)$ are insensitive to this choice, and we use
$\gamma^{\infty} = 2/3$. The fitting procedure is illustrated in 
Figure \ref{nu_gamma_plot}, where we show the fit for $\nu_0(V)$, the 
log moment and the resulting $\gamma_0(V)$. We did the fits with
$\nu_0(V_0)$, $\gamma_0(V_0)$, and $q_0(V_0)$ as free parameters, and with
$\gamma_0(V_0)$ constrained to the experimental value of 2.97. (This
experimental value was determined so that the thermal expansion for
the subsequent EOS overlies the recommended curve of Touloukian\cite{touloukian}
from 100 to 1200 K.) The unconstrained fit gives $\gamma_0(V_0) = 2.88$.
Constraining $\gamma_0(V_0)$ increases the RMS error of  the fit from 
$4.9 \times 10^{-2}$ to $5.7 \times 10^{-2}$ THz, which is not a
significant increase. Also, we note that constraining $\gamma_0(V_0)$
does not change $\gamma_0(V)$ at smaller $V$. For this reason,
we believe that the constrained fit is the best overall approximation for
$\gamma^{\rm ion}(V)$ up to twofold compression. This is shown as the solid
curve in Figure \ref{nu_gamma_plot}. The corresponding parameters
are
$\nu_0(V_0)  =  3.46$ THz,
$\gamma_0(V_0)  =  2.97$,
$q_0(V_0)  =  1.37$.
We do not show the unconstrained fit in Figure \ref{nu_gamma_plot} because
it is too close to the constrained fit. The dot-dashed curves correspond
to $q={\rm const}$, $\gamma(V)=\gamma(V_0)(V/V_0)^q$. The value 
$q=1$ gives a $\gamma$ that is too high over this density range, 
while $q=1.7$\cite{heinz} is too low.  Jamieson {\it et al.}\cite{jamieson}
used $q=1$ and an ambient value $\gamma(V_0) = 3.215$, which results
in the dashed curve. We believe that their $\gamma$ is too high at all
volumes considered here. Any constant value of $q$ will
lead to a $\gamma$ that is too small at very high compression. 

\section{Equation of State}

The complete EOS is determined when the Helmholtz free energy $F$ is given
as a function of volume and temperature. In this section we describe our 
form for $F$ and give the numerical parameters for fcc Au.
First we write the free energy as in Eq.~(\ref{f_total}) as the sum of
the static lattice energy $\phi_0$, an ion motion free energy
$F^{\rm ion}$, and an electronic excitation term $F^{\rm el}$.
For the ion motion free energy we use the quasi-harmonic approximation,
and further specialize to the high temperature Debye model. Thus
we take the ion free energy to be given by Eqs.~(\ref{f_debye}) and
(\ref{debye_fn}), where we 
identify $\omega_D$ with the log moment $\omega_0$ of the phonon frequencies.
As discussed in section II, this gives the correct classical limit, and
leads to extremely small errors at low $T$ compared to the full phonon
spectrum. This way we are able to give a closed form expression for $F^{\rm ion}$,
which is very convenient in numerical applications. The volume dependence
of $\omega_0$ is that given  in Eq.~(\ref{nu_of_v}) with parameters
given above. For the final EOS we have replaced
the fitted value of $\omega_0(V_0)$ with the experimental one. This is
essentially a small shift in the absolute entropy, which has negligible
effect on the results.

 Electronic excitations give a small contribution to the thermodynamics, which
we approximate by
\begin{equation}
F^{\rm el}(V,T) = -\frac{1}{2} N \Gamma(V) T^2 \, .
\label{f_el}
\end{equation}
The Sommerfeld coefficient $\Gamma$ is proportional to the electronic density of
states at the Fermi energy, $\Gamma = (\pi^2/3) k_B^2 g(\epsilon_f)$.
Our calculations show that $d \ln g(\epsilon_f) / d \ln V \approx 0.76$
over the range of densities considered, so we take
\begin{equation}
\Gamma(V) = \Gamma(V_0) \left(V/V_0 \right)^\kappa
\label{dos_of_v}
\end{equation}
with
$ \kappa = 0.76$.
The calculations give $\Gamma(V_0) = 6.7 \times 10^{-4}$ J/mol K$^2$, while
the measured value from the low-temperature specific heat is\cite{hultgren}
$\Gamma(V_0) = 7.28 \times 10^{-4}$ J/mol K$^2$. The measured low-temperature
specific heat coefficient $\Gamma_{\rm expt}$ is expected to be enhanced 
with respect to the bare value $\Gamma$ as a result of
electron-phonon interactions  by a factor $(1 + \lambda)$, where $\lambda$
is the dimensionless electron-phonon mass enhancement parameter. 
The value for gold is $\lambda \approx 0.05 - 0.15$,\cite{allen} 
which agrees with
our calculated $\Gamma/\Gamma_{\rm expt} = 1.09$. At high temperatures,
$k_B T > \hbar \omega_0$, the phonon mass enhancement becomes ineffective. 
Hence we use our calculated density of states in the remainder of this paper.

The static lattice energy is needed to complete the free energy. We adopt 
the functional form due to Vinet {\it et al.}\cite{vinet}
\begin{eqnarray}
\phi_0(V) & = &  \frac{4 V^* B^*}{(B_1^* -1)^2}
             \left[ 1 - (1+X) e^{X}\right]
               \nonumber \\
       X & = & \frac{3}{2} (B_1^* -1)
         \left[\left(\frac{V}{V^*}\right)^{1/3} -1 \right]
\label{phi0}
\end{eqnarray}
which is parameterized by the volume at the minimum, $V^*$, the bulk
modulus, $B^*$ and its pressure derivative, $B_1^*$. These parameters
have been determined empirically by requiring that the EOS reproduce the
ambient volume, and ultrasonic data for $B_S$ at ambient conditions.\cite{simmons}
For the latter, we adopt the value $B_S = 173$ GPa. Ultrasonic data
gives values for $\left( \partial B_S/ \partial P \right)_T$ from
5.2 to 6.4. Therefore, to complete the EOS, we require that it
reproduce the measured slope of the Hugoniot, which we take
from the fit\cite{llnl} $U_s = 3.12 {\rm km/s} + 1.521 U_p$
relating the shock velocity $U_s$ to the particle velocity $U_p$.
The resulting parameters give
$\left( \partial B_S/ \partial P \right)_T = 5.49$, consistent 
with ultrasonic data. 

The complete set of parameters for the free energy
of fcc Au are summarized here. The  static lattice energy is given by
Eq.~(\ref{phi0}) with
\begin{eqnarray}
V^*  & = & 10.0834 \; {\rm cm}^3/{\rm mol}
\nonumber \\
B^*  & = & 180.0 \; {\rm GPa}
\nonumber \\
B_1^* & = & 5.55 \, .
\end{eqnarray}
The ion motion free energy is given by Eqs.~(\ref{f_debye}) and (\ref{debye_fn})
with the volume dependence of $\omega_D$ given by Eq.~(\ref{nu_of_v}) 
(with $\nu_0 \rightarrow \omega_D$, $\gamma_0 \rightarrow \gamma$, etc.). 
The parameters $A$ and $B$ are given in terms of $\gamma(V_0)$ and
$q(V_0)$ by Eq.~(\ref{AB_gammaq}). The numerical values are
\begin{eqnarray}
V_0 & = & 10.212 \; {\rm cm}^3/{\rm mol}
\nonumber \\
\omega_D(V_0) & = & 22.9 \times 10^{12} \; {\rm s}^{-1}
\nonumber \\
\gamma(V_0) & = & 2.97
\nonumber \\
q(V_0) & = & 1.3677
\nonumber \\
\gamma^{\infty} & = & 2/3 \, .
\end{eqnarray}
Finally, the electronic excitation free energy is given by Eqs.~(\ref{f_el})
and (\ref{dos_of_v}) with
\begin{eqnarray}
\Gamma(V_0) & = & 6.7 \times 10^{-4} \; {\rm J/mol \, K}^2
\nonumber \\
\kappa & = & 0.76 \, .
\end{eqnarray}
The following calculations use these parameters for the free energy of
fcc Au. Once the free energy is known, the pressure, internal energy,
etc., can be evaluated. We calculate the Hugoniot by fixing a value of
the volume and solving for the temperature such that the jump condition,
$E - E_0 = \frac{1}{2} (P + P_0) (V_0 - V)$ ,
is solved. Here $E_0$, $V_0$ and $P_0$ correspond to the initial state, taken
to be ambient temperature and pressure.

\begin{figure}
\noindent
\includegraphics[scale=0.45]{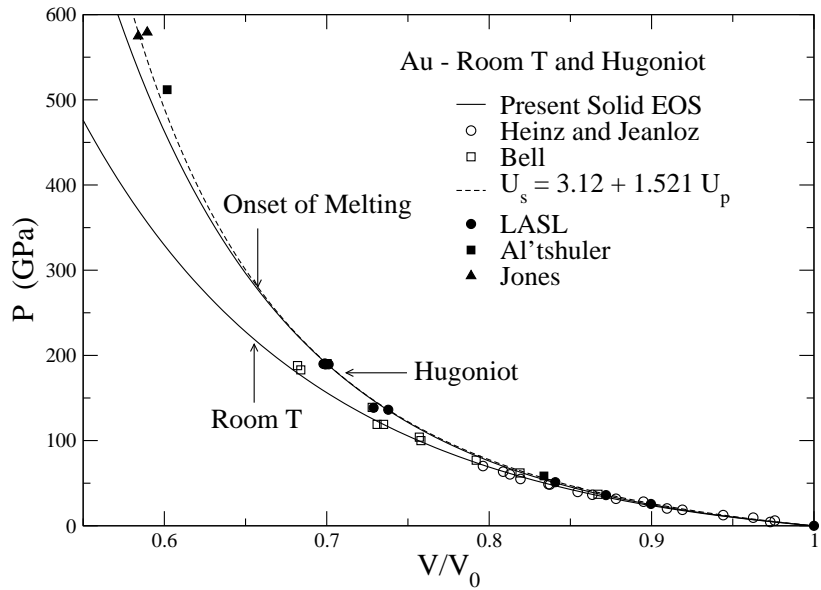}
\caption{Hugoniot and room temperature isotherm of Au. Solid curves are present 
        solid EOS. Solid symbols are Hugoniot data.\cite{lasl,altshuler,jones}
        Note that there are five overlapping Hugoniot points at $V/V_0 \approx 0.7$.
        Open symbols are room temperature data.\cite{heinz,bell_sccm} 
        For the Bell {\it et al.} data,\cite{bell_sccm} pressures are from
        the extrapolated ruby standard. The onset of shock melting is estimated
        at $P \approx 280$ GPa, as marked.}
\label{hug_roomt_plot}
\end{figure}

Figure \ref{hug_roomt_plot} shows the Hugoniot and room temperature
isotherm for the present EOS. The solid symbols are the Hugoniot 
data.\cite{lasl,altshuler,jones} The dashed line corresponds to the
linear fit for $U_s(U_p)$.
 Also shown as the open circles are the
room temperature data from Heinz and Jeanloz\cite{heinz} taken to 70 GPa
with the ruby pressure scale. The open squares are the room temperature
data of Bell {\it et al.}\cite{bell_sccm} to 188 GPa. The pressure
shown in the graph is taken from the ruby scale, which is extrapolated
from a lower pressure calibration.\cite{mbss} Our calculations agree
with these data to 3\% at the highest pressure, indicating support for
the extrapolation of the ruby scale. 

For $V > 0.656 V_0$, the Hugoniot lies in the solid. By matching the 
measured Hugoniot with our theoretically based Gr\"uneisen parameter,
we confirm the accuracy of our room temperature isotherm, and 
we have high confidence in our EOS to this density. Combining our
calculated Hugoniot temperatures with a Lindemann melting curve,
we estimate that the Hugoniot enters a solid-liquid coexistence
region at $V/V_0 =0.656$, $P = 280$~GPa. 
A more extensive discussion of shock melting is given in the following section.
There we find that explicitly accounting for melting leads to good agreement
with the high pressure Hugoniot data, indicating that our solid
EOS is valid to densities of $V/V_0 \approx 0.6$.

\begin{figure}
\noindent
\includegraphics[scale=0.45]{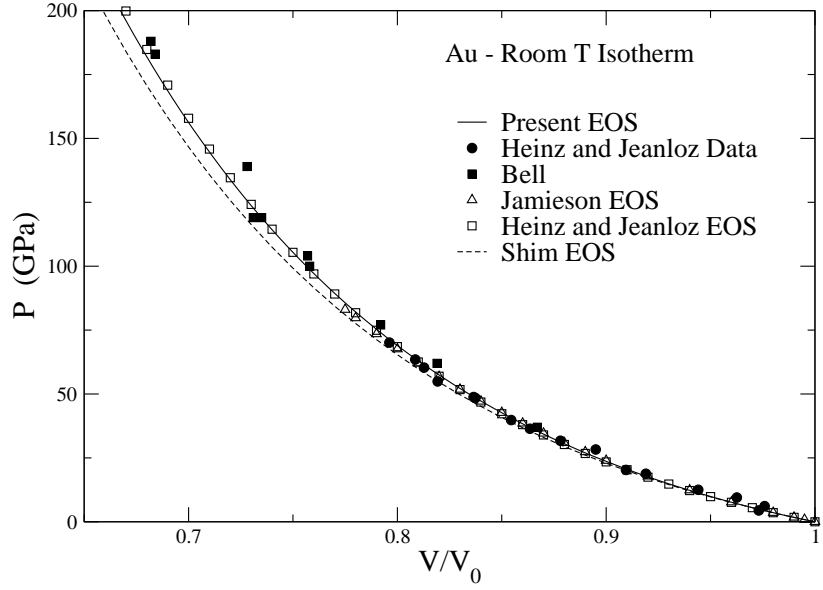}
\caption{Room temperature isotherm of Au. Solid curve is present EOS.
         Solid symbols are data.\cite{heinz,bell_sccm} Open triangles
         are the EOS of Jamieson {\it et al.}\cite{jamieson} Open
         squares are EOS of Heinz and Jeanloz.\cite{heinz} Dashed curve
         is EOS of Shim {\it et al.}\cite{shim}}
\label{room_t_plot}
\end{figure}

Figure \ref{room_t_plot} shows our room temperature isotherm along with data from
Heinz and Jeanloz,\cite{heinz} and Bell {\it et al.}\cite{bell_sccm}  Also
shown are some of the proposed EOS standards. Jamieson {\it et al.}\cite{jamieson}
used the Hugoniot as a reference, and calculated the room temperature isotherm
using a Mie-Gr\"uneisen EOS. They assumed $\gamma/V = {\rm const}$,
and used $\gamma(V_0) = 3.215$, so their $\gamma$ is always larger than
ours. This results in lower room temperature pressures than ours, but 
by restricting
their analysis to $V/V_0 > 0.775$, the impact of the assumed $\gamma$ 
is minimized. The Heinz and Jeanloz EOS is based on extrapolating their
room temperature data, with some consideration of the shock data. Our
room temperature isotherm is in good agreement with Heinz and Jeanloz
to 200 GPa. The recently proposed EOS of Shim et al.\cite{shim} is significantly
lower than all the other standards. It is lower than the present analysis
by  10 GPa at $V/V_0=0.7$, corresponding to $P=156$ GPa.

\begin{figure}
\noindent
\includegraphics[scale=0.45]{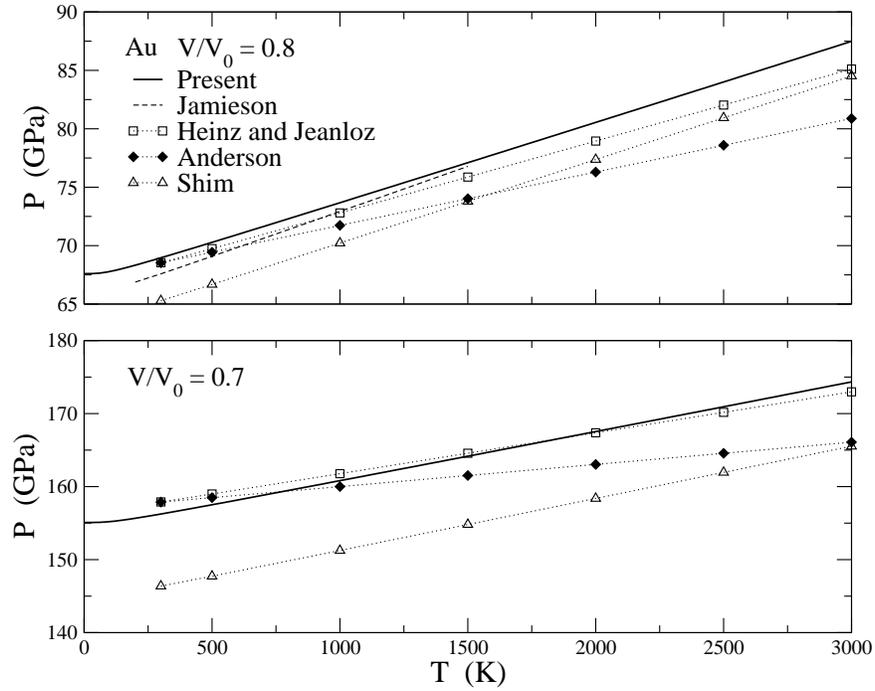}
\caption{Isochores of Au. Solid curves are the present EOS. Dashed curve
         is Jamieson {\it et al.}\cite{jamieson} Open squares are Heinz and
         Jeanloz.\cite{heinz} Filled diamonds are 
         Anderson {\it et al.}\cite{anderson_au} Open triangles are Shim
         {\it et al.}\cite{shim}.}
\label{isochores_plot}
\end{figure}

Figure \ref{isochores_plot} shows $P(T)$ along two isochores in 
comparison with various equations of state. At $V/V_0 = 0.8$,
Jamieson {\it et al.} give somewhat lower $P$ at 300 K with a larger $dP/dT$,
in keeping with their higher $\gamma$. Heinz and Jeanloz\cite{heinz} are in the
best overall agreement with the present pressures, with somewhat lower
$dP/dT$. Shim {\it et al.}\cite{shim} give a somewhat higher $dP/dT$ 
with $P$ generally low due to their large offset at room temperature. 
Anderson {\it et al.}\cite{anderson_au} adopted the room temperature isotherm
from Heinz and Jeanloz.\cite{heinz} Their EOS gives $dP/dT$ substantially
too low at compressed volumes.

We have tabulated $P(V,T)$ in table \ref{au_eos_table} in the same
format adopted by other authors. As discussed, we have extended the range of
densities to $V/V_0 = 0.6$ and the temperatures to 5000 K, except 
where such temperatures are thought to lie in the liquid phase. 
Boettger\cite{boettger_au} has proposed an extension of the Au
300 K standard to 500 GPa based on LDA calculations. He gives $P=344$ GPa
at $V/V_0 = 0.6$ compared to 329 GPa from our semi-empirical EOS.
Given the scatter in the high pressure Hugoniot data, and our approximate
treatment of melting, this 4.6\% difference is probably within the
uncertainties of the present analysis at this high density.

\begin{table}
\caption{Tabulated Pressures for Au. Compression $\eta = 1 - V/V_{\rm 300}$
         where $V_{\rm 300}$ is the volume at $T = 300$~K and $P= 0$.
         Pressures in GPa. Values in parenthesis are the first liquid states
          for each $\eta$,
         included for interpolation. Remaining liquid states left blank.}
\label{au_eos_table}
\begin{tabular}{rrrrrrrrrrrr}
\hline
$\eta$  &  300 K   &    500 K &    1000 K &   1500 K  &   2000 K  &   2500 K  &   3000 K & 3500 K & 4000 K & 4500 K & 5000 K \\
\hline
\hline
   0.00  & 0.00 & 1.44 &    5.08 &    8.73 &   (12.40) &   &   &   &    &    &   \\
   0.02  & 3.57 & 5.00 &    8.61 &   12.24 &   (15.88) &    &   &   &    &    &    \\
   0.04  & 7.65 & 9.07 &   12.65 &   16.25 &   19.87 &   (23.50) &    &    &    &    &    \\
  0.06  & 12.31 & 13.71 &   17.27 &   20.85 &   24.44 &   (28.04) &    &    &    &    &    \\
  0.08  & 17.61 & 19.01 &   22.53 &   26.09 &   29.66 &   33.24 &   (36.83) &    &    &    &    \\
  0.10  & 23.65 & 25.03 &   28.54 &   32.07 &   35.61 &   39.17 &   42.74 &   (46.33) &    &    &    \\
  0.12  & 30.53 & 31.90 &   35.38 &   38.88 &   42.41 &   45.94 &   49.50 &   (53.06) &    &    &    \\
  0.14  & 38.36 & 39.71 &   43.17 &   46.65 &   50.16 &   53.67 &   57.20 &   60.74 &   (64.30) &    &    \\
  0.16  & 47.27 & 48.61 &   52.04 &   55.51 &   58.99 &   62.48 &   65.99 &   69.51 &   73.05 &   (76.60) &    \\
  0.18  & 57.41 & 58.74 &   62.15 &   65.60 &   69.06 &   72.53 &   76.02 &   79.53 &   83.04 &   (86.57) &    \\
  0.20  & 68.96 & 70.28 &   73.67 &   77.09 &   80.54 &   83.99 &   87.47 &   90.95 &   94.45 &   97.96 &  (101.49) \\
  0.22  & 82.12 & 83.43 &   86.80 &   90.21 &   93.63 &   97.07 &  100.53 &  104.00 &  107.48 &  110.98 &  114.49 \\
  0.24  & 97.13 & 98.43 &  101.78 &  105.17 &  108.58 &  112.00 &  115.44 &  118.90 &  122.37 &  125.85 &  129.34 \\
 0.26  & 114.27 & 115.55 &  118.89 &  122.26 &  125.65 &  129.07 &  132.49 &  135.93 &  139.39 &  142.85 &  146.34 \\
 0.28  & 133.85 & 135.12 &  138.44 &  141.80 &  145.18 &  148.58 &  151.99 &  155.42 &  158.86 &  162.32 &  165.79 \\
 0.30  & 156.25 & 157.51 &  160.81 &  164.16 &  167.53 &  170.92 &  174.32 &  177.74 &  181.17 &  184.62 &  188.08 \\
 0.32  & 181.92 & 183.17 &  186.46 &  189.79 &  193.16 &  196.53 &  199.93 &  203.34 &  206.76 &  210.20 &  213.65 \\
 0.34  & 211.38 & 212.63 &  215.90 &  219.22 &  222.58 &  225.95 &  229.34 &  232.74 &  236.16 &  239.59 &  243.04 \\
 0.36  & 245.26 & 246.49 &  249.75 &  253.07 &  256.42 &  259.79 &  263.17 &  266.57 &  269.99 &  273.41 &  276.86 \\
 0.38  & 284.30 & 285.52 &  288.76 &  292.08 &  295.43 &  298.79 &  302.17 &  305.57 &  308.98 &  312.41 &  315.85 \\
 0.40  & 329.37 & 330.58 &  333.82 &  337.14 &  340.48 &  343.85 &  347.23 &  350.63 &  354.04 &  357.47 &  360.91 \\
\hline
\end{tabular}
\end{table}

\section{Melting and the Hugoniot}

\begin{figure}
\noindent
\includegraphics[scale=0.45]{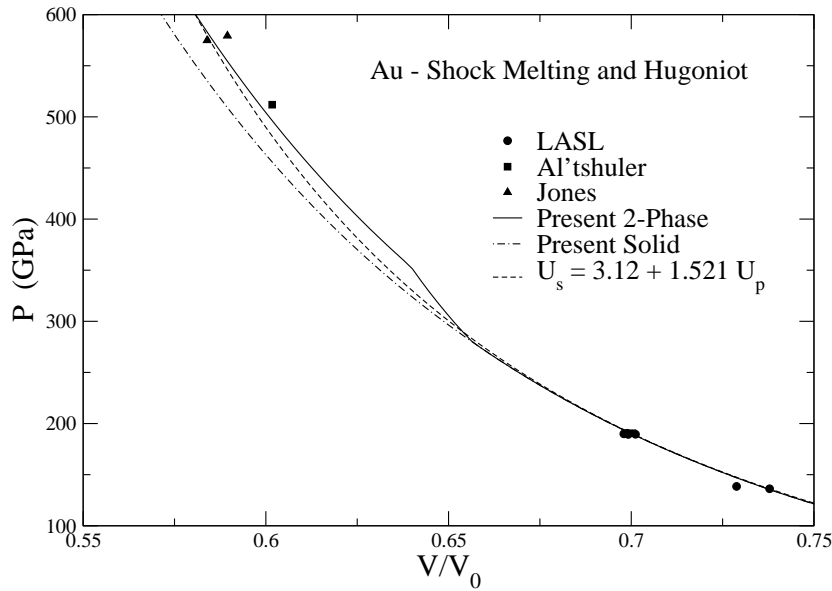}
\caption{Hugoniot of Au, including the effects of melting. Solid curve
         is present two-phase (solid and liquid) EOS. Dot-dashed curve is
         solid EOS only. Dashed curve is linear fit to $U_s(U_p)$ 
         data.\cite{llnl} Solid symbols are measured 
         points.\cite{lasl,altshuler,jones}}
\label{two_phase_hug}
\end{figure}

In order to investigate the effects of melting on the Hugoniot, we have 
constructed a two-phase model free energy. The ion free energy in the 
liquid is based on the assumptions that $C_V^{\rm ion} = 3Nk_B$, 
which is reasonable for temperatures near melting, and that 
$\Delta S_V^{\rm ion}$, the entropy difference between solid and liquid 
at fixed volume is $0.8 N k_B$. These are empirically based model
assumptions.\cite{grover} A statistical mechanical basis for these observations
is discussed by Wallace,\cite{wallace_ldyn} who argues for the universality of
$\Delta S_V^{\rm ion}$. Beyond these assumptions, we require the energy 
of the liquid with respect to the solid to fully determine the liquid 
free energy. We do this by imposing that the melting curve, obtained by
equating the pressures and Gibbs free energies of liquid and solid, follow 
the Lindemann rule in the form
\begin{equation}
\frac{T_m}{\omega_0^2(V_s) V_s^{2/3}} = {\rm const} \, ,
\end{equation}
where $V_s$ is the volume on the solidus.
The electron excitation free energy is assumed to be the same in the
liquid as in the solid. Application of this method to Cu leads to a shock
melting threshold  of 228 GPa, which compares well with 232 GPa obtained
by Hayes {\it et al.}\cite{hayes_cu} by analyzing sound speed data. The Au
Hugoniot has been calculated for
the two-phase model allowing for coexistence 
in the shocked state.\cite{greeff_ti}
Figure \ref{two_phase_hug} shows the resulting Hugoniot in the pressure-volume
plane. Also shown in the figure are the data points and the linear $U_s(U_p)$
fit. The dot-dashed curve is the Hugoniot for the solid only. The
boundaries of the coexistence region are visible as kinks on the solid
curve at 280 and 350 GPa. Above 350 GPa, the Hugoniot is in pure liquid phase.
A similar anomaly is shown for Al in a two-phase calculation by 
Chisolm {\it et al.}\cite{chisolm_al} A significant enhancement of the
Hugoniot pressure of Au due to melting was also shown in calculations by
Godwal {\it et al.}\cite{godwal}, however they predict a much lower
shock melting threshold than we do.
Just above complete melting, the liquid 
Hugoniot lies above the solid by 30 GPa. The linear $U_s(U_p)$ curve
goes smoothly through the melting region to intersect the liquid data.

Our two-phase EOS gives good agreement with the high pressure data, which
suggests that our cold energy is valid to compressions of 
$V/V_0 \approx 0.6$. Electronic excitations have practically no effect on
the $P(V)$ Hugoniot in the solid phase, although they significantly affect
the temperature. At higher temperatures in the
liquid, they act to soften the Hugoniot by absorbing energy.
Coincidentally, neglecting both electronic excitations and melting gives a
Hugoniot that agrees with the linear $U_s(U_p)$ curve quite well to
650 GPa. This is an accidental cancellation of errors. Ignoring melting 
and electronic excitations gives a temperature that is too high by
$\sim 10^4$~K at this pressure. This offsets the neglect of the pressure
enhancement due to melting. It is not recommended to ignore
either melting or electronic excitations in this high temperature regime.

\section{Conclusions}

By combining {\it ab initio} calculations of elastic moduli and
zone boundary phonons with interpolations based on a force constant model,
we have calculated moments of the phonon frequencies of fcc Au to
twofold compression. This allows us to calculate the associated Gr\"uneisen
parameters. In particular, we have focused on $\gamma_0$ which corresponds
to the classical limit. We emphasize that the classical limit dominates
the thermal pressure in the EOS. We find that the frequently used
form $\gamma(V) = \gamma(V_0) (V/V_0)^q$ does not represent the volume
dependence of $\gamma$ well. We have used an expansion to $2^{\rm nd}$ order
in $V$ with a physical asymptotic limit, which fits the calculated moments
well and is consistent with the measured value of $\gamma$ at ambient pressure
and temperature.
Using the theoretical $\gamma(V)$ and electron excitation free energy,
we have constructed a semi-empirical free energy for the solid, which we believe
to be as accurate as possible.  This free energy is given as a parameterized
closed form expression and the resulting $P(V,T)$ is given in tabular
form to $V/V_0 =0.6$.

Our static lattice energy is empirical, and is verified by comparison
with the measured Hugoniot, so that our EOS can be regarded as a
generalization from the Hugoniot with a physically based $\gamma$.
In the solid phase, electronic excitations have a small effect.
This, together with the dominance of the classical limit in the
vibrational free energy, means that the widely used 
Mie-Gr\"uneisen approximation,
that $\left( \partial P /\partial E \right)_V$ is independent of $T$,
is accurate.
At the highest compressions the physics affecting the Hugoniot becomes 
more complicated. Melting is predicted to begin at 280 GPa 
and $V/V_0 = 0.656$. Complete melting is estimated to lead 
to a 30 GPa increase in the Hugoniot pressure over the solid. 
At these densities and
higher, the Hugoniot temperature is rising rapidly and electronic excitations 
are playing an increasing role. The highest Hugoniot data are at
580 GPa, where the temperature is calculated to be above $2 \times 10^4$ K.
The Mie-Gr\"uneisen approximation is not expected to be valid for 
interpolating between the Hugoniot and room temperature at this high
compression,
because of the strong effects of melting and electronic excitations.
These effects need to be taken explicitly into account, as has been done here.

Regarding EOS standards, our analysis gives a room temperature
$P(V)$ curve that agrees well with that of Heinz and Jeanloz\cite{heinz}
to 200 GPa. Our
EOS gives $\left( \partial P /\partial T \right)_V$ generally somewhat larger than
theirs, and we believe that in this regard our EOS is preferred.
The EOS of Jamieson {\it et al.}\cite{jamieson} is based on reduction
of shock data, and was originally given to 80 GPa. It gives somewhat
lower pressures than ours, which is due to their use of a Gr\"uneisen
parameter that is too large at all volumes. Extrapolation of their
room temperature isotherm to higher pressures\cite{akahama} is not recommended.
Anderson {\it et al.}\cite{anderson_au} adopted the room temperature
isotherm of Heinz and Jeanloz\cite{heinz}, while giving a different
thermal pressure. Their EOS gives $\left( \partial P /\partial T\right)_V$
which is substantially too low under compression, and is not recommended for
high temperatures. Simultaneous compression of Au and Pt showed\cite{akahama}
that the Pt standard of Holmes {\it et al.}\cite{holmes_pt} gave a
pressure higher by 15 and 20 GPa than the Au standards of Heinz 
and Jeanloz\cite{heinz} and Jamieson,\cite{jamieson} respectively,
at a pressure of 150 GPa.
While the extrapolated Jamieson isotherm is expected to be somewhat low,
our analysis agrees with the room temperature
isotherm of Heinz and Jeanloz, suggesting that the discrepancy between
the Au and Pt pressures is
due to errors in the Pt standard. Given the importance of accurate
pressure standards, the status of the Pt EOS seems to warrant
further investigation.

\begin{acknowledgments}

We thank J. C. Boettger, J. D. Johnson, E. D. Chisolm and S. Crockett
for helpful discussions. This work was supported by the U. S. Department of Energy
under Contract No. W-7405-ENG-36.

\end{acknowledgments}


\end{document}